\documentstyle[11pt,epsf]{article}

\newcommand{\bea}{\begin{eqnarray}}
\newcommand{\eea}{\end{eqnarray}}

\def\void{}
\def\labelmark{}
\newenvironment{formula}[1]{\def\labelname{#1}
\ifx\void\labelname\def\junk{\begin{displaymath}}
\else\def\junk{\begin{equation}\label{\labelname}}\fi\junk}%
{\ifx\void\labelname\def\junk{\end{displaymath}}
\else\def\junk{\end{equation}}\fi\junk\labelmark\def\labelname{}}
{\ifx\void\labelname\def\junk{\end{array}\end{displaymath}}
\else\def\junk{\end{array}\right.\end{equation}}
\fi\junk\labelmark\def\labelname{}\def\junk{}
}
\newcommand{\beq}{\begin{formula}}
\newcommand{\eeq}{\end{formula}}
\newcommand{\beqv}{\begin{formula}{}}
\newtheorem{guess}{Conjecture}

\begin{document}

\begin{titlepage}

\null
\begin{flushright}
SU-4240-627\\
\end{flushright}
\vspace{20mm}

\begin{center}
\bf\Large Singular Vertices and the \\
          Triangulation Space of the $D-$sphere
\end{center}

\vspace{5mm}

\begin{center}
{\bf S. Catterall, G. Thorleifsson}\\
Physics Department, Syracuse University,\\
Syracuse, NY 13244.\\
{\bf J. Kogut}\\
Loomis Laboratory, University of Illinois at Urbana,\\
1110 W. Green St, Urbana, IL 61801-3080.\\
{\bf R. Renken}\\
Department of Physics, University of Central Florida,\\
Orlando, FL 32816.
\end{center}

\begin{center}
\today
\end{center}

\vspace{10mm}
      
\begin{abstract}
By a sequence of numerical experiments we demonstrate that generic
triangulations of the $D-$sphere for $D>3$ contain one {\it singular}
$(D-3)-$simplex.  The mean number of elementary $D-$simplices sharing
this simplex increases with the volume of the triangulation
according to a simple power law. The lower dimension
subsimplices associated with this $(D-3)-$simplex also show a 
singular behaviour. Possible consequences for the DT model of
four-dimensional quantum gravity are discussed.
\end{abstract}

\vfill
\end{titlepage}

\section{Introduction}

It has been well established that two dimensional quantum gravity 
can be recovered as the scaling limit of models of random triangulations,
see for example \cite{generic}. Performing a
sum over such simplicial manifolds generates the correct
integral over physically inequivalent metrics. As a natural
extension of these ideas it has been proposed that triangulations
of higher dimensional manifolds can form the basis of a general
regularization scheme for gravity \cite{prop}. 
In general, such simplicial manifolds are constructed by gluing together
$D-$dimensional simplices across their $(D-1)-$dimensional faces
so as to form a closed triangulation with some fixed topology. Additional
manifold restrictions are commonly imposed to ensure that all
simplices consist of a set of $(D+1)$ distinct labels and every
subsimplex is unique.

The ansatz for the partition function in general dimension $D$ 
then takes the form
\begin{equation}
Z \; = \; \sum_{V}\; e^{-\kappa V}\Omega_D\left(V\right),
\end{equation}
where $\kappa$ is a bare cosmological constant conjugate to
the volume or total number of $D-$simplices $V$. 
The {\it microcanonical} partition function
$\Omega_D\left(V\right)$ is given by 
\begin{equation}
\Omega_D\left(V\right)=\sum_{T\left(V\right)}\; e^{-S\left(T,\kappa_i\right)}
\label{z}.
\end{equation}

The sum over triangulations $T$ is confined to those with
volume $V$, with a weight determined by a discrete
action $S\left(\kappa_i\right)$ governed by a set of couplings 
$\{\kappa_i\}$. In the case of four dimensions the simplest action
contains only one such coupling $\kappa_0$ which can be identified with
the bare (inverse) Newton constant. The corresponding analog of the 
Einstein-Hilbert action can be taken to be the total number of
vertices $N_0$ in the triangulation.

Current interest in this model stems from the results of
numerical simulations which have revealed a non-trivial phase
structure in four dimensions. Between a crumpled phase with
large negative curvature at small
$\kappa_0$ and an elongated, branched polymer phase  at large $\kappa_0$
there is evidence of a continuous phase transition. The existence of
such a {\it non-perturbative} critical point offers the possibility of
a continuum limit describing quantum gravity \cite{lots}. However,
it would be fair to say that the nature of this continuum theory
is only just beginning to be explored. 

One of the major
problems impeding progress in this direction has been 
the lack of any analytic methods for handling the sum over
four-dimensional simplicial geometries. The structure of
the triangulation space and its implications for the measure over
simplicial geometries are unknown. In this paper we hope to
make some progress in this direction using numerical simulation to
identify a class of
triangulations which dominate the microcanonical partition
function $\Omega_D\left(V\right)$ for large volumes $V$. 

This work is motivated by a recent observation
that typical triangulations in the crumpled phase of
the four-dimensional model are characterized by
$2$ highly degenerate or singular vertices \cite{japanese}. 
Singular vertices are vertices that are shared by a large number of
simplices -- a number which diverges linearly with the total volume
of the triangulation. 

In Section 2 we describe our conjecture for the structure of
the dominant $D-$dimensional triangulations together with
supporting numerical results. Section 3 makes plausible why  
such configurations might be entropically favoured and uses a
simple geometrical model to explain some of the observed volume
dependencies. 
Section 4 contains a discussion of
singular vertex dynamics and its practical implications for
numerical simulations.
Finally Section 5 contains a brief discussion
of possible consequences of this singular structure. Specifically
we discuss the issue of
an exponential bound in four dimensions. 

\section{Structure of the Triangulation Space}

Our Monte Carlo simulations employ a set of local, ergodic and
topology preserving moves\footnote{For a
dimension independent implementation see \cite{simon}.}
(see for example \cite{lots,gross}). 
We have set the
action $S$ to zero so that the simulations explore equally all
triangulations contributing to the partition function Eq.~\ref{z} 

Let us define the {\it local volume} associated with an $i-$simplex
as the number of $D-$simplices containing that $i-$simplex.
We then say that the $i-$simplex is {\it singular} if its local
volume diverges with the total volume $V$ (total number
of $D-$simplices).
Our results can then be summarized in a simple conjecture:

\begin{guess}
The function $\Omega_D\left(V\right)$ is saturated as $V\to\infty$
by triangulations which contain one {\it singular} $(D-3)-$simplex.
\end{guess} 

\begin{figure}
\centering
\epsfxsize=4.7in \epsfbox{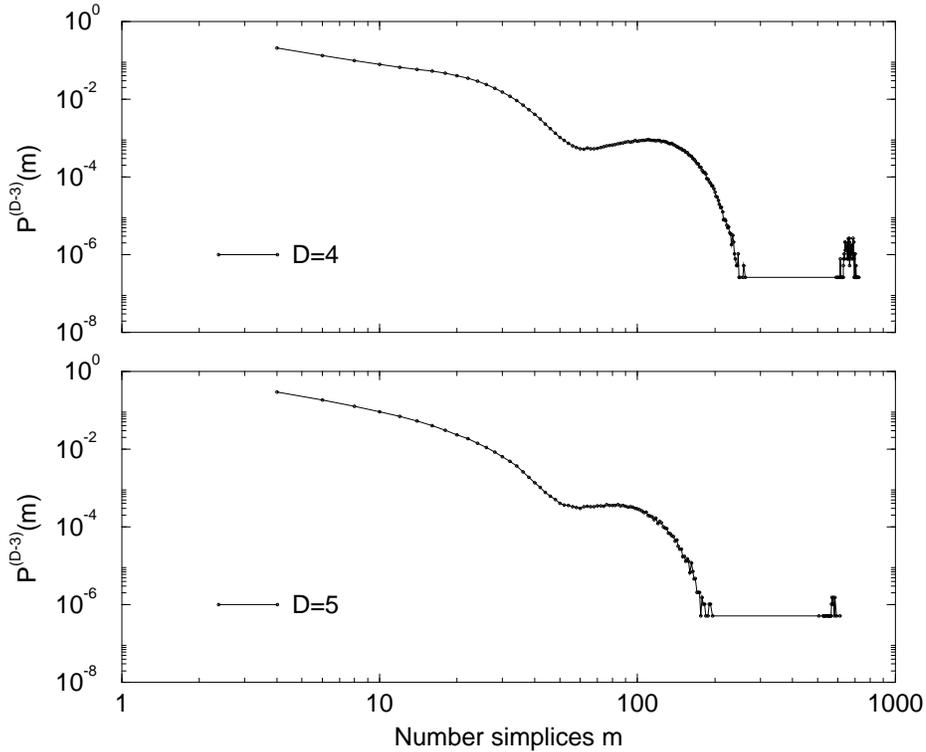}
\caption{The normalized distribution of 
local volumes for $(D-3)-$simplices in four
and five dimensions.} 
\label{dist}
\end{figure}

This is illustrated by Fig.\ \ref{dist} which shows
the (normalized) distribution of $(D-3)-$simplices with a
given local volume $m$, denoted by $P^{\left(D-3\right)}\left(m\right)$. 
The two plots correspond to 
four and five dimensions 
where the
singular object is a link and a triangle respectively. The data in both
cases
come from simulations in which the total
volume $V=32000$.
Clearly, both distributions possess an isolated peak in the tail
corresponding to $(D-3)-$simplices which are
common to a large number of $D-$simplices. Furthermore, we have 
observed that this peak corresponds to the presence 
{\it for each triangulation} of precisely one such singular $(D-3)-$simplex.

\begin{figure}
\centering
\epsfxsize=4.7in \epsfbox{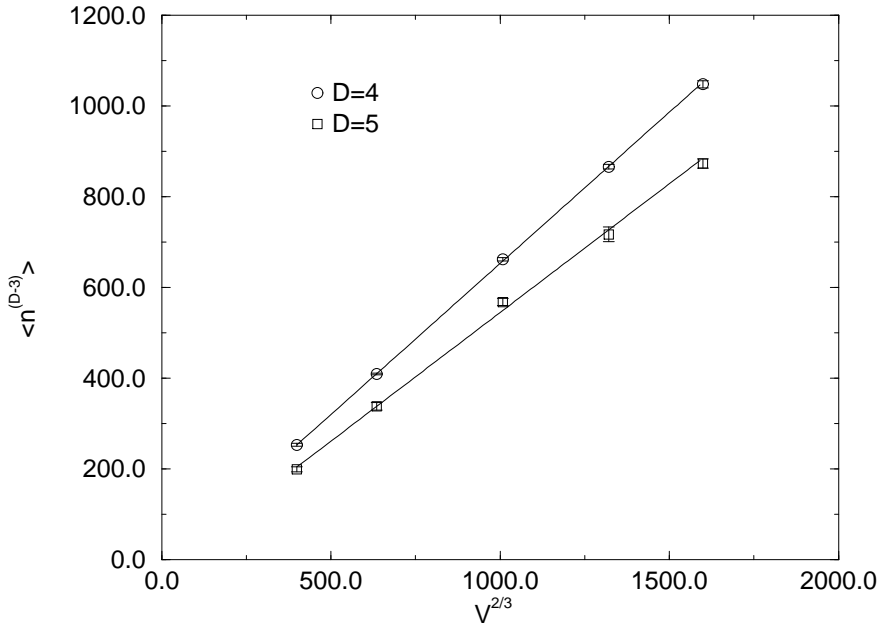}
\caption{The mean local volume of the $(D-3)-$simplex 
vs the total volume $V$ for
four and five dimensions. Note that we plot the data as a function
of $V^{\frac{2}{3}}$.} 
\label{spower}
\end{figure}

Fig.~\ \ref{spower} shows the scaling of the 
mean local volume of this singular $(D-3)-$simplex
with the total triangulation volume,
again for $D=4$ and 5. We
have utilized lattices of size $V=8000$ to $V=64000$.
For large volumes it can be seen that
these results support the notion of a power-law
divergence. Furthermore, the data is 
consistent with a unique, simple power growth
given by the solid lines. These  correspond to
a power of $\frac{2}{3}$. The justification
for the choice of this power will be discussed in Section 3.

\begin{figure}
\centering
\epsfxsize=4.0in \epsfbox{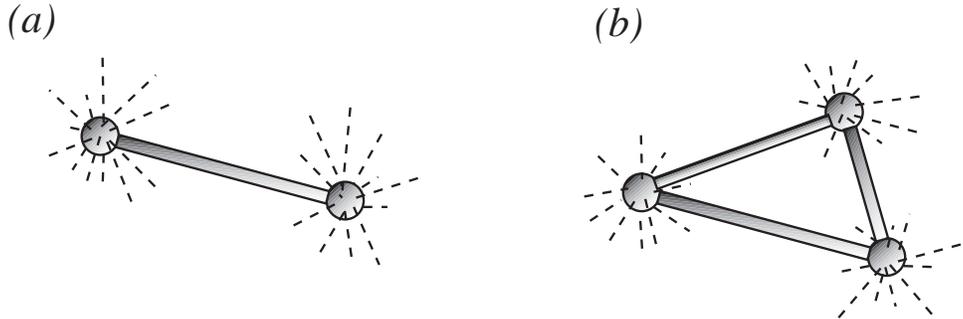}
\caption{The singular structure in a) $D=4$ b) $D=5$. The balls
correspond to singular vertices which overlap along singular links and
triangles. }
\label{pic1}
\end{figure}

Associated with this {\it primary} $(D-3)-$simplex is a cascade of other
singular simplices corresponding to all its possible subsimplices. 
Thus we observe precisely 
\begin{center}
$\left(\begin{array}{c}D-2\\i+1\end{array}\right)$ 
\end{center}
{\it secondary singular} $i-$simplices, where
$i=0,\ldots,(D-4)$, whose
local volumes diverge in the thermodynamic
limit. Thus in four dimensions we see exactly two singular
vertices corresponding to the endpoints of the original
singular link. In five dimensions we have one singular triangle,
three singular links corresponding to its edges and three singular
vertices. Fig.~\ \ref{pic1} illustrates this for dimensions four (a)
and five (b).
This pattern continues in higher dimension, for
example in six dimensions the dominant triangulations
have four singular vertices. 

\begin{figure}
\centering
\epsfxsize=4.7in \epsfbox{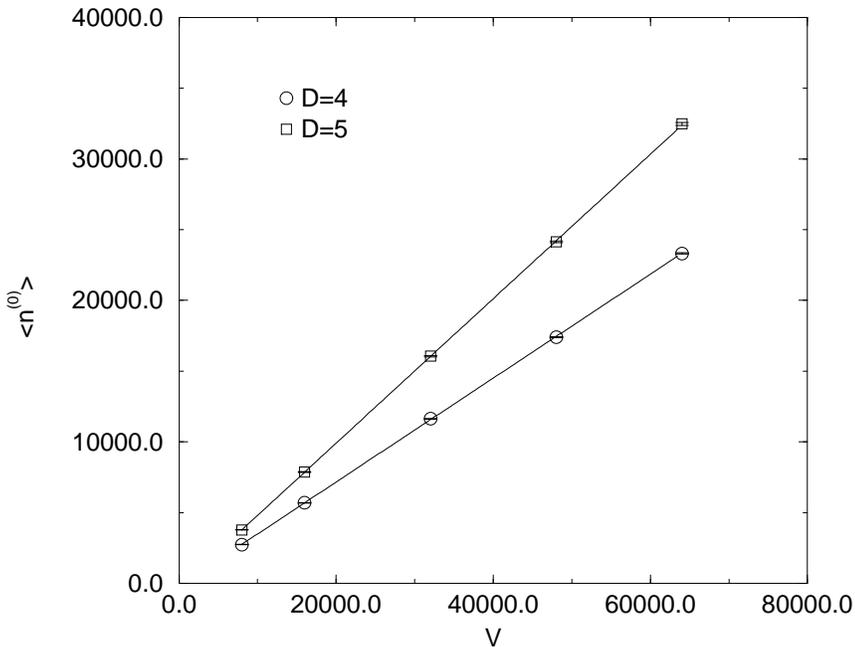}
\caption{The mean local volume of singular vertices vs total volume $V$ for
four and five dimensions.}
\label{nodes}
\end{figure}

\begin{figure}
\centering
\epsfxsize=4.7in \epsfbox{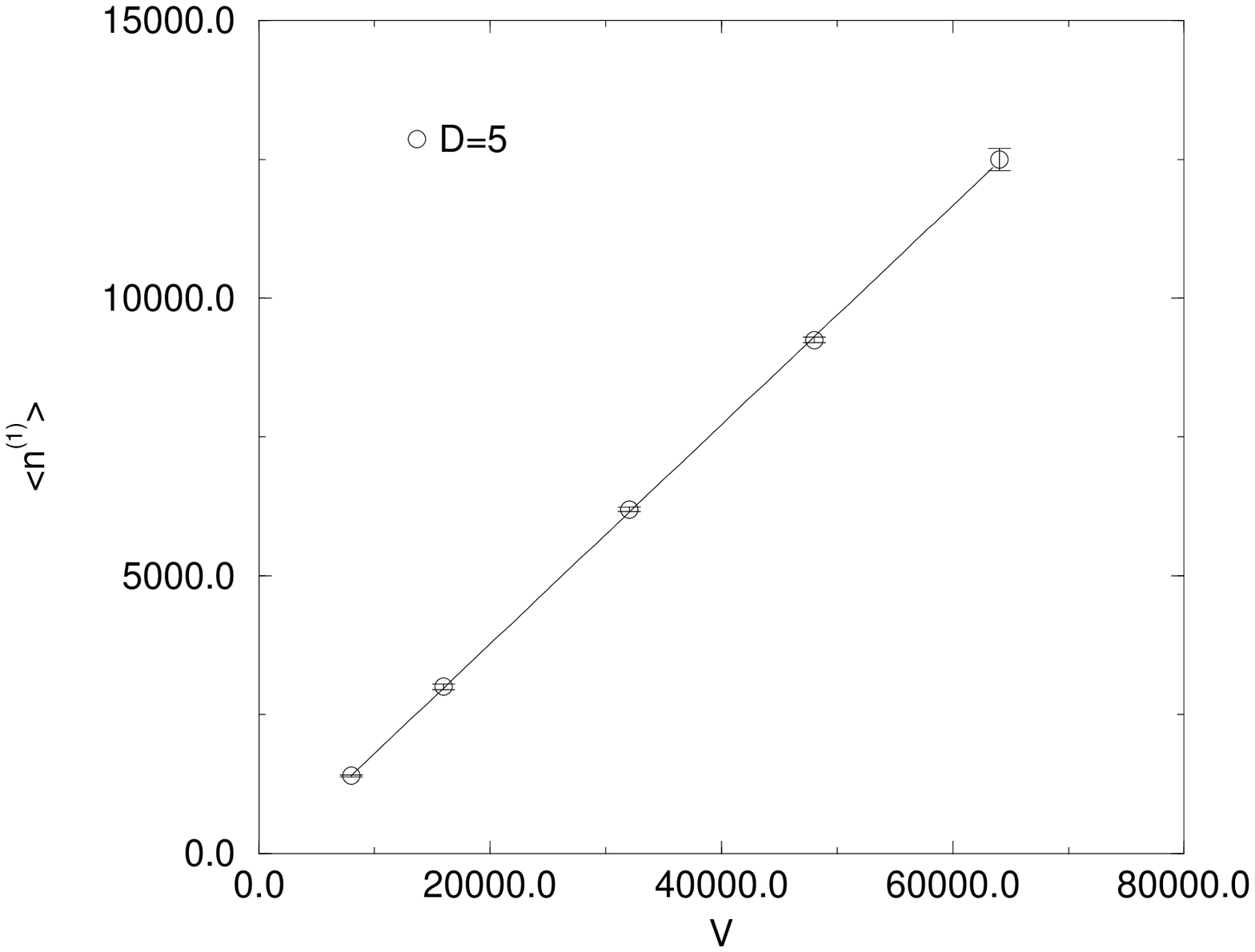}
\caption{The mean local volume of singular links vs total volume $V$ for
five dimensions.}
\label{links}
\end{figure}

In contrast to the primary singular simplex we find that the
mean local volumes of these secondary
singular simplices grow {\it linearly} with the
volume of the triangulation. Fig.~\ \ref{nodes}
shows a plot of the mean singular vertex volume for both four
and five dimensions. 
The linear growth of the singular
link volume in five dimensions is shown in Fig.~\ \ref{links}.
We have observed that
each triangulation is symmetric with respect
to exchange of two singular simplices of a given
dimension - they have approximately the
same local volumes.
On the basis of this numerical evidence we can state
another hypothesis.

\begin{guess}
While the primary singular $(D-3)-$simplex has a local volume which
grows as some power $\;p\sim \frac{2}{3}\;$ of the total volume, the secondary
singular simplices have local volumes which grow linearly with
volume.
\end{guess}


We have also recorded the mean number of $D-$simplices $V_{\rm ns}$
which are
{\it not} associated with any of the singular vertices. The number of
these again increases linearly with volume $V_{\rm ns}=c_{\rm ns}\,V$. 
Since the total volume
is fixed at $V$ there is a relationship between the local volumes
$\omega_i=c_i\,V$ of singular $i-$simplices and the non-singular
simplices $c_{\rm ns}\, V$. 
In the infinite volume limit (where the $(D-3)-$simplex
does not contribute) it is straightforward to verify that the
following relation holds between the coefficients $c_i$. 

\begin{equation}
1-c_{\rm ns}=\sum_{i=0}^{D-4}\left(\begin{array}
{c}D-2\\i+1\end{array}\right)
\left(-1\right)^i c_i\;.
\label{prove}
\end{equation}

In four dimensions the measured $c_{\rm ns}=0.279(2)$ which is to 
be compared with its value computed from the above 
relation, $c_{\rm ns}=0.266(3)$. Given
the systematic errors present in these fits we regard this
as quite satisfactory agreement. Notice that this is completely
consistent with the observation that the singular link
volume increases sublinearly as $V\to\infty$. If that were
not the case the righthand side of Eq.~\ref{prove} would receive
another contribution from the links.
It is also satisfied in five dimensions where the measured 
value $c_{\rm ns}=0.045(1)$ is
statistically consistent with the value estimated using this
relation, $c_{\rm ns}=0.058(6)$.

\section{Entropy Considerations}

Given these conjectures about the nature of the configuration
space, is it possible to understand why this very special class
of triangulations is entropically favored?  Why for instance are
there no singular $(D-2)-$simplices?  In this section we give
some heuristic arguments for this, and also try to explain the
nature of the observed power-law divergence of the
singular $(D-3)-$simplices.

Consider the local volume associated with a particular
$i-$simplex. It is composed
of a set of $D-$simplices each of which contains the $(i+1)$ vertex
labels of the $i-$simplex in question. Take the set of vertex labels
associated to these local volume $D-$simplices and remove the
$(i+1)$ labels of the common $i-$simplex. The remaining vertex labels
then constitute a triangulation
of a $(D-i-1)-$sphere. This sphere is the boundary of the dual to the
simplex - a $(D-i)-$dimensional volume. The volume of this sphere is
just proportional to the original $i-$simplex local volume. 

\begin{figure}
\centering
\epsfxsize=4.0in \epsfbox{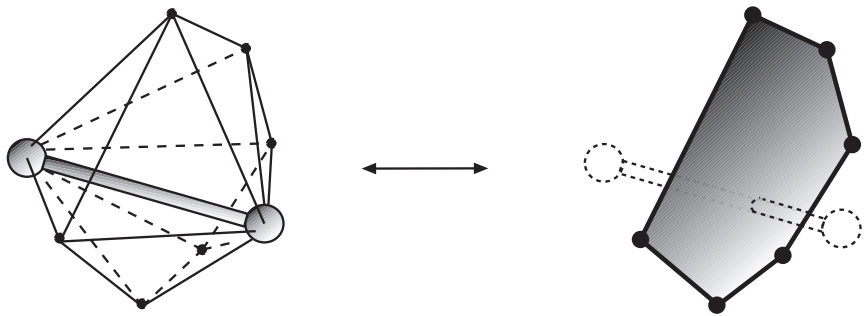}
\caption{The dual area and its bounding triangulated circle for a link
in three dimensions.}
\label{linkpic}
\end{figure}

For example, in three dimensions, the 
dual to a link is an area element whose boundary
is a triangulation of the circle.  
The vertices defining this circle are just those
obtained from the simplices making up
the local volume of the link excluding the endpoints of
the link itself.  By definition, the link's local volume is then just
proportional to the number of vertices on the circle. This is
illustrated in Fig.~\ \ref{linkpic} which shows the simplices making
up the local volume of a link in three dimensions, its dual area and
the associated $1-$sphere - the triangulated circle.

We can now ascribe
a local entropy to an $i-$simplex with local volume 
$\omega_i$ by counting the number of ways
of gluing together the $\omega_i$ simplices containing it.  Each of these
gluings corresponds to a distinct triangulation of the associated
dual $(D-i-1)-$sphere.
This allows us to map the problem of
counting the number of ways of achieving a certain
local volume by gluing together $D-$simplices into
the enumeration of all the 
possible triangulations of the dual $(D-i-1)-$sphere.
Specifically, the local entropy of
the $i-$simplex with local volume $\omega_i$ is just
determined by the number of triangulations of the
associated dual $(D-i-1)-$sphere with volume $\omega_i$.   

\begin{figure}
\centering
\epsfxsize=4.7in \epsfbox{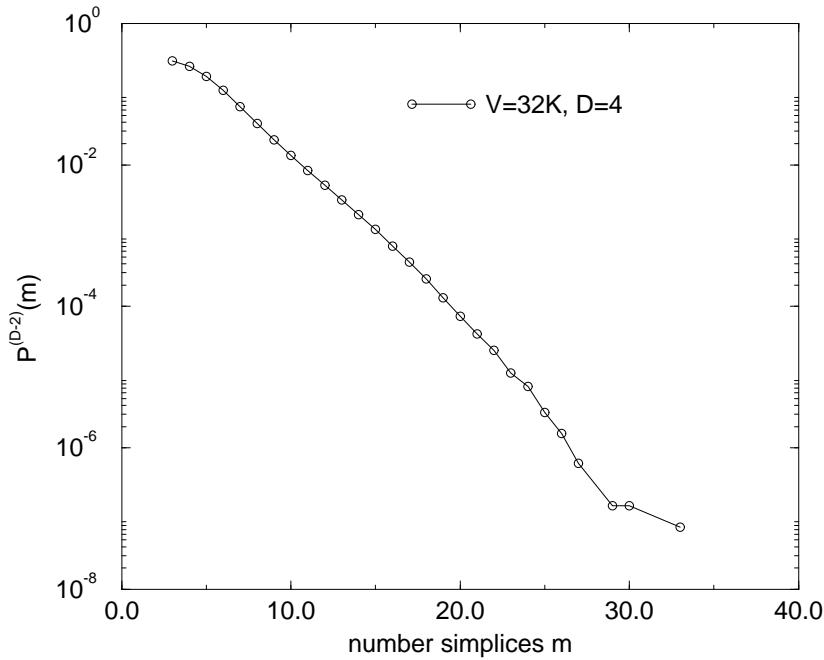}
\caption{The normalized distribution of local volumes for
$(D-2)-$simplices in four dimensions.}
\label{curv}
\end{figure}

For $(D-2)-$simplices the relevant sphere is 
$S^1$. There is only one distinct way of
arranging the simplices in its local volume. Equivalently,
there is a unique triangulation of $S^1$ for any local
volume $\omega_{D-2}$. Thus the local entropy of a $(D-2)-$simplex does
not increase as its local volume increases. It is {\it not}
entropically favoured for such a $(D-2)-$simplex to acquire a large
local volume. 
Indeed, 
the number of $(D-2)-$simplices possessing large local
volumes falls off (approximately) exponentially fast. 
This is seen in Fig.~\ \ref{curv} which shows the (normalized) distribution of 
$(D-2)-$simplices common to $m$ $D-$simplices, denoted by $P^{\left(D-2\right)}
\left(m\right)$.
The data corresponds to
four dimensions but similar results are obtained in dimensions
three through six. 
Notice that the curvature density is associated to $(D-2)-$simplices and
hence never receives any singular contributions.

The situation is very different for
$(D-3)-$simplices. The local entropy associated to
such an object is again given by the number of ways of gluing
together the simplices
constituting its local volume. By our arguments this
corresponds exactly to the number of
triangulations of the sphere $S^2$ with area equal to the
local volume $\omega_{D-3}$.
This grows exponentially with 
the local volume $\omega_{D-3}$.

Similarly, for
$i-$simplices with $i=D-4,\ldots,0$ the local entropy is
determined by the number of triangulations of the sphere 
$S^{D-i-1}=S^3, S^4,\ldots,S^{D-1}$
containing $\omega_i$ faces. This is known to
increase {\it at least} as fast as exponentially with local volume
$\omega_i$. 
Thus, in contrast to $(D-2)-$simplices, 
simplices of dimension $i=0,\ldots,(D-3)$  can 
maximize their local entropy by acquiring large local
volumes.

If we start out with some arbitrary 
triangulation of fixed volume and perform a random set of
local moves it is clear that individual $i-$simplices with
$i=0,\ldots,D-3$ are unstable
to growing their local volumes. We can imagine 
qualitatively that individual simplices {\it compete} with
each other subject to the constraint that the topology and
total number of
simplices remains fixed. 
Is it possible to understand why a {\it single} $(D-3)-$simplex will eventually
dominate?  While we cannot construct a rigorous argument that
this should necessarily be so the following line of
reasoning renders it, we believe, at least plausible.

Suppose we have a configuration with  some number $n$,
not necessarily $(D-2)$, 
potential singular vertices. If $n$ is smaller than $(D-2)$ the system
can increase its entropy by acquiring more singular vertices. What 
stops the number $n$ growing arbitrarily? The entropy of each
vertex increases with local volume; thus these vertices will want to grow
their local volumes as large as possible. Ultimately, this means
that the potential singular vertices will
want to get as close as possible to each other
so that they can share simplices. This overlapping of local
volumes will be maximal when the potential singular vertices
form the vertices of some simplex $S$. The overlaps between
vertex volumes are then associated with subsimplices of the simplex $S$.
These subsimplices too can gain entropy by becoming singular - their local
volumes acquiring a finite fraction of all the simplices in
the triangulation.

Thus the simplex
$S$ which results from the intersection of such singular simplices
will itself become singular. But we have seen that singular
simplices of dimension
$(D-2)$ and greater are not entropically favoured. 
Thus the degeneration process stops when
$S$ has dimension of $(D-3)$ - it becomes the primary, singular
$(D-3)-$simplex. Then the number of
singular vertices cannot increase beyond $(D-2)$.
This qualitative argument is able to account
for at least the local stability of the configurations that are
seen in the numerical simulations. It makes credible the notion
that these configurations are at least {\it local} maxima of
the total entropy.

It is also possible to understand the origin of the power-law
divergence of the $(D-3)-$simplex. The dual $2-$sphere associated with
this simplex is the boundary of the overlap of two $3-$spheres
dual to the secondary $(D-4)-$simplices. If we assume that
the simplices associated with these $3-$spheres are uniformly
distributed 
over the surface of the spheres, then simple
geometry allows us to compute the number on the boundary of
the overlap -- the
local volume $\omega_{D-3}$. 

Introduce a length scale or radius
for the $3-$sphere by equating the classical volume
formula for a $3-$sphere with the number of $D-$simplices in the
local volume of a $(D-4)-$simplex -- $\omega_{D-4}=c_{D-4}\; V$.
Uniform distribution of simplices implies
that the radius of the $2-$sphere is linearly related to the
radius of this $3-$sphere. 
Using this we can obtain a prediction for the volume of
the $2-$sphere or equivalently the $(D-3)-$simplex local
volume $\omega_{D-3}$:  
\begin{equation}
\omega_{D-3}=3\pi\left(\frac{c_{D-4}}{2\pi^2}\right)^{\frac{2}{3}}\,
V^{\frac{2}{3}}.
\label{vol}
\end{equation}
This equation predicts both the volume dependence and
coefficient $c_{D-3}$ of the singular simplex divergence (using
as input the measured coefficient $c_{D-4}$). We have
already seen that a $\frac{2}{3}$ power of the volume
is very consistent with the observed scaling of $\omega_{D-3}$.
In four dimensions,
the predicted value of $c_{D-3}=0.661(2)$ which compares very
well with its value estimated from a last squares fit to
the $(D-3)-$simplex data, $c_{D-3}=0.667(5)$.
In five dimensions $c_{D-3}=0.437(2)$
from Eq.\ \ref{vol},
while our best (using the largest three
volumes) fit estimate is $c_{D-3}=0.49(2)$. 

These quantitative tests lend strong support to our basic
geometrical picture. Essentially these triangulations are
formed by taking $D-2$ singular vertices with approximately equal
volumes $c_{0}V$ and gluing them together to form
a singular $(D-3)-$simplex. A large fraction of all the $D-$simplices
have then been used to create this special $D-$ball. The remainder
are used to help glue faces of this $D-$ball together in order to
create a triangulation with the correct $S^D$ topology. In the
context of the crumpled phase of DT gravity this
basic structure forms a non-perturbative background about which
small fluctuations in triangulation occur.

\section{Singular vertex dynamics}

To what extent can we trust the results of these numerical
experiments? Is it possible that these configurations are not
truly dominant but act as local stationary points of the entropy
which trap the configurations and effectively break the
ergodicity of the algorithms? We have tried to address these
questions by performing a number of tests.

Current algorithms used in Monte Carlo simulation of these models
rely on a sequence of $D+1$ local moves or
re-triangulations which are known to be ergodic on the
full space of triangulations $T$ (at least for $D<5$ \cite{gross}).
However, to approach the continuum limit in a regular fashion
lattice simulations are restricted to the microcanonical
ensemble $T_V$ characterized by $\Omega_D\left(V\right)$. Actually to
allow the elementary moves to be carried out it is necessary that
the volume be allowed to fluctuate by at least $+/-D$.
Typical simulation strategies have relaxed this restraint still
further and allowed the volume to fluctuate about the target volume
by some amount $\Delta V>D$.

Unfortunately, the ergodic properties of the algorithm when
thus restricted to a fixed volume slice $T_V$ are unknown.
One simple scenario might be that the system possesses
`volume barriers' $B\left(V\right)$ of all sizes up to some
volume dependent limit, 
\begin{equation}
B\left(V\right)\le B_{\rm max}\left(V\right)\;.
\end{equation}
We might then expect a practical breakdown of ergodicity if the allowed
fluctuation volume $\Delta V$ becomes less than $B_{\rm
max}\left(V\right)$. It is possible that such an effect might
be important in effectively trapping configurations in the vicinity
of one of these singular triangulations.  We have investigated this
issue by performing simulations with a variety of $\Delta V$. In order to
keep control of the systematic error
associated with the finite volume $V$ we have chosen to take
measurements only when the volume of the triangulation lies within some
distance $\delta$ of the target volume $V$.
In practice we have set $\delta=10$. A breaking of ergodicity would
be signaled by a dependence of expectation values on $\Delta V$.

Table 1 summarizes our results in the
case of $D=4$ for a volume $V=4000$ (similar results have been obtained
in $D=3$ and at other lattice sizes). We 
observe no statistically significant
dependence in the mean vertex number $\left<N_0\right>$
and mean intrinsic extent $\left<L\right>$ on $\Delta V$ over
a wide range in $\Delta V$.
This is a very encouraging and, in principle, non-trivial result. It
is in agreement with earlier studies \cite{aergod,bergod} which have
have not shown any evidence of ergodicity breaking in four and five
dimensions.

\begin{table}
\begin{center}
\begin{tabular}{|l|l|r|}\hline
$\Delta V$&  $\left<L\right>$&  $\left<N_0\right>$ \\ \hline
6         &  9.168(5)        & 226.7(2)          \\ 
10        &  9.156(6)        & 226.1(7)          \\ 
14        &  9.187(14)       & 226.9(7)          \\ 
20        &  9.153(20)       & 226.1(7)          \\ 
32        &  9.157(10)       & 226.6(3)          \\ 
45        &  9.137(19)       & 226.7(8)          \\ 
63        &  9.156(15)       & 226.6(8)          \\ 
100       &  9.165(6)        & 226.5(1)          \\ 
200       &  9.166(9)        & 226.3(4)          \\ \hline
\end{tabular}
\caption{The dependence of expectation values on fluctuation parameter
$\Delta V$ for $D=4$ and $V=4000$.}
\end{center}
\end{table}

\begin{figure}
\centering
\epsfxsize=4.7in
\epsfbox{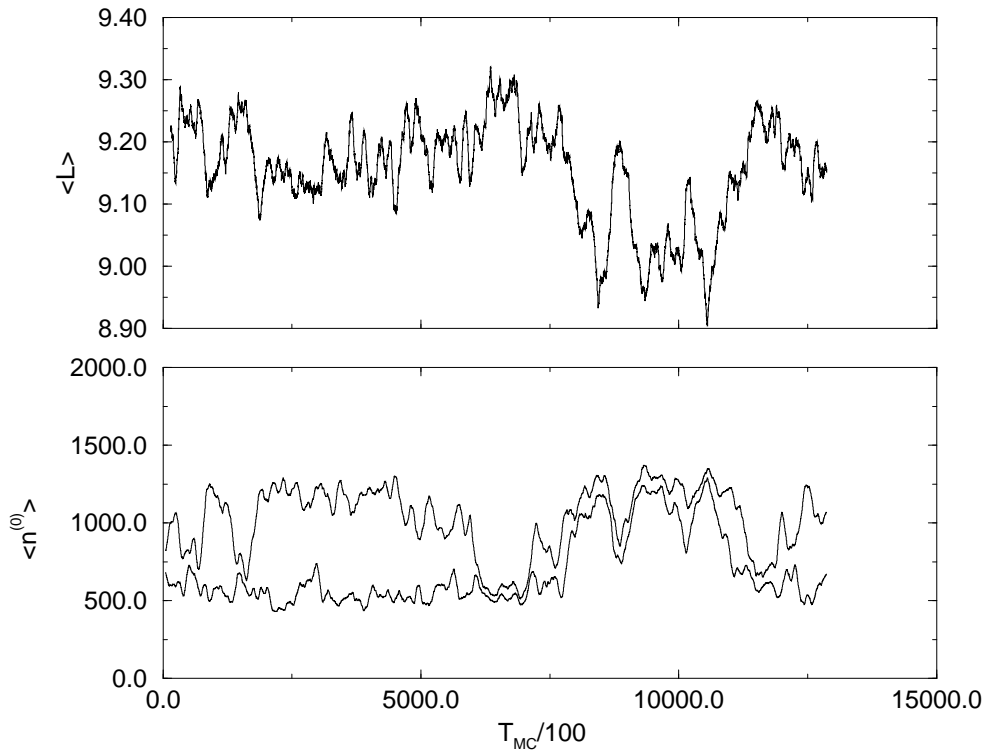}
\caption{The MC time series for the mean intrinsic extent 
$\left\langle L\right\rangle$ (top) and the local volumes
associated to the two most singular vertices (bottom) for $V=4000$
and $D=4$.}
\label{lseries}
\end{figure}

However, we have
observed {\it very} long autocorrelation times in both
observables which can easily obscure this result if only
short runs are employed. The upper graph of
Fig.~\ \ref{lseries} illustrates this with
a plot of the Monte Carlo time series for the mean 
extent $\left<L\right>$
for a lattice of size $V=4000$ with $\Delta V=10$. We can see that the
system makes occasional excursions to `super-crumpled' states
with small extent and remains trapped there for many tens of
thousands of sweeps before it can tunnel back. The typical
timescale between such events is of order one million sweeps!\footnote{
One sweep corresponds to $V$ attempted local moves.}

We have observed the same
problems over wide ranges in the
fluctuation volume $\Delta V$.
The frequency of such large fluctuation events seems independent of
this parameter. The lower graph of Fig.~\ \ref{lseries} contains
a plot of the two vertices with the largest local
volumes for the same Monte Carlo history. It is clear that
these rare fluctuation events are associated to the appearance and
disappearance of singular vertices. For small volumes it appears that
the system can sometimes have zero or a single singular
vertex - contrary
to the claims made in the previous section which state that
configurations with two such vertices are dominant. 
However, that claim is true only for
$V\to\infty$ and it is clear that for small volumes tunneling
between distinct free energy minima (labeled by differing numbers
of singular vertices) can occur. However, our simulations revealed
no evidence that the tunneling time depends on $\Delta V$. 

\begin{figure}
\centering
\epsfxsize=4.7in
\epsfbox{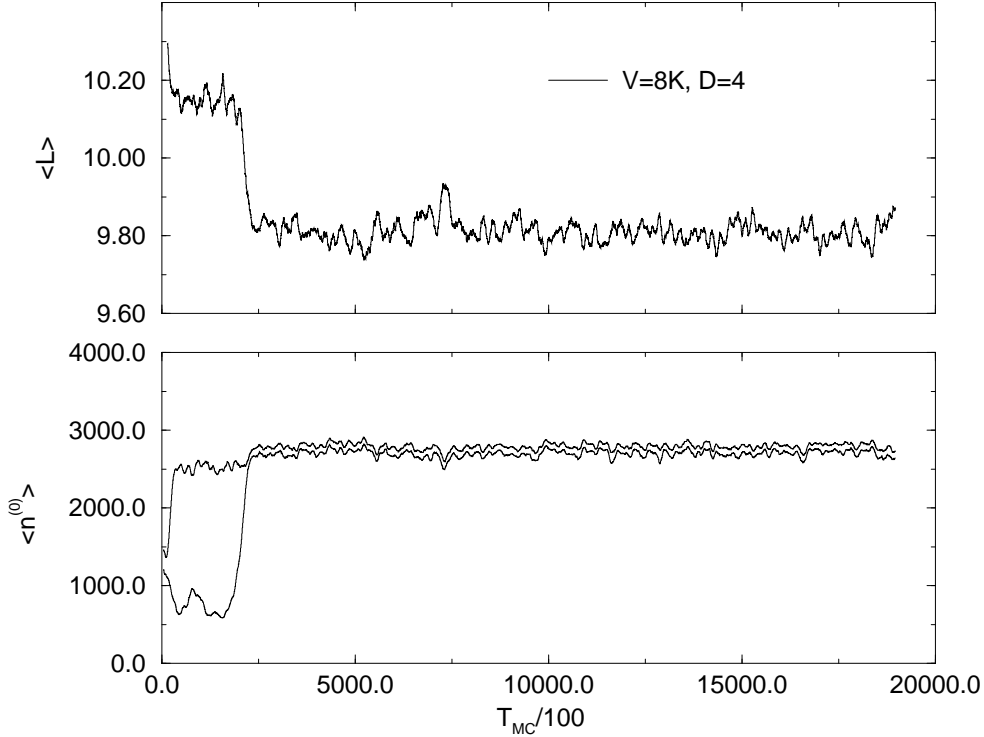}
\caption{ The same as the previous figure for $V=8000$. }
\label{lseries2}
\end{figure}

We have also observed long transient effects in trying to equilibrate 
larger volumes. Fig.~\ \ref{lseries2}
illustrates a typical run for $V=8K$ in
four dimensions.  We show both $\left<L\right>$ and the two most
singular vertices.
It is clear
that the system appears to settle down into an
equilibrium state with small fluctuations after perhaps a few
tens of thousands of sweeps. This state 
contains precisely one singular vertex. However it is clearly 
metastable and after a further few hundred thousand sweeps 
undergoes a rapid transition to a more crumpled state 
possessing two singular vertices. 
We have not managed to observe any
subsequent reverse tunneling. This transient behaviour has
been observed for many different fluctuation volumes $\Delta V$ and
a variety of initial state configurations. Similar behaviour has
also been seen in five and six dimensions at small to
intermediate volumes. It is also consistent with recent findings
reported by Hotta et al \cite{japanese} who observe a relaxation to
two singular vertices in four dimensions independent of start
configuration.

In three dimensions we observe no singular vertices and no
corresponding tunneling or metastable behaviour. In this sense
three dimensions appears qualitatively different from
four and higher dimensions.

To summarize this section. We have looked for evidence that the 
singular states are metastable as a consequence of
ergodicity breaking in the simulation algorithm. Such a breaking
would be signaled by a dependence of expectation values on
the fluctuation parameter $\Delta V$. Over a wide range of this
parameter we observe no such dependence. While we observe
long autocorrelation times at small volume associated with
singular vertex dynamics, this behaviour appears to disappear
for large volume and we are led to conclude that the singular
states do indeed saturate $\Omega_D\left(V\right)$ in the
thermodynamic limit. 

\section{Possible consequences}

Our numerical results imply that the 
typical triangulations as $V\to\infty$ are singular configurations -
they consist of a set of $D-2$ singular vertices assembled into
a singular $(D-3)-$simplex. The local volume associated
to the $(D-3)-$simplex increases as a fractional power $p\sim \frac{2}{3}$
of the total volume. In contrast, the local volume associated to its secondary
subsimplices increases in proportion to the volume. We have
argued in section 3 that this structure is at least a local
maximum of the entropy function for triangulations with
fixed volume. The numerical results of section 4 imply that it
appears to be a global maximum.
The question arises whether this structural information can 
be used to cast light on a variety of other issues in DT
gravity.

Specifically, in four dimensions can we learn anything
about the possibility of an exponential bound in the microcanonical
partition function; i.e.\
$\Omega_4\left(V\right)\sim
\exp{\mu V}$? Such a bound is needed to take the thermodynamic limit.
A proof for triangulated manifolds has so far eluded all efforts
(although a related proof
for metric ball coverings does exist \cite{ball}).  Direct
numerical estimates of $\Omega_4\left(V\right)$, while
consistent with such a bound, are unfortunately
plagued with large finite size effects which require rather
delicate analysis \cite{usbound,ambbound,enzbound,baby}.

In four dimensions the 
important simplicial manifolds consist of two
elementary $4-$balls containing the singular vertices joined 
along a common link.  Approximately two thirds of the 
total volume is locked up in these balls, which become
independent in the $V \to \infty$ limit.  

Thus the triangulation of the four-sphere contains within it
two independent $3-$sphere boundaries carrying a large fraction of the
total volume. 
Provided the number of triangulations
of the $3-$sphere is bounded exponentially (which is believed
to be true from previous numerical simulations \cite{3d}), the
proof of the $4D$ bound rests on showing that 
the number of ways these balls
can be glued together, using the remaining one third of the
volume, is exponentially bounded.  This seems to be 
an easier task than to show that the triangulation space
is exponentially bounded for arbitrary triangulations.
However we have not been able to prove this or its
obvious generalizations to higher dimensions.

Indeed, there is one question
which we do not understand concerning this structure.
Why does the primary singular $(D-3)-$simplex only diverge sublinearly
with volume in contrast to the linear divergence of all
lower dimension singular simplices? Does this signal a different
volume behaviour of the entropy function for $S^2$ as compared with
$S^r,\,r=3,\ldots$ or is it a simple consequence of the constraints
which are present? Further work, both analytic and numerical, is
needed to understand the consequences of this and
related features of the structure presented here.

\vfill

\noindent
{\bf Acknowledgments}
 
The calculations reported here were supported, in part, by grants
NSF PHY-9503371, PHY-9200148 and from research funds
provided by Syracuse University. Some computations were performed 
using facilities at the NorthEast Parallel Architectures Center, NPAC.
We would like to thanks Varghese John for a careful reading of
the manuscript.

\vfill


\vfill

\begin{thebibliography}{99}
\bibitem{generic}
 F. David, {\it Simplicial Quantum Gravity and Random Lattices},
  (hep-th/9303127), Lectures given at Les Houches Summer School 
  on Gravitation and Quantization, Session LVII, 
  Les Houches, France, 1992;  \\
 J. Ambj\o rn, {\it Quantization of Geometry}, (hep-th/9411179), 
  Lectures given at Les Houches Summer School on Fluctuating 
  Geometries and Statistical Mechanics, Session LXII, 
  Les Houches, France 1994;  \\
 P. Di Francesco, P. Ginzparg and J. Zinn-Justin, Phys. Rep. 
  254 (1995) 1.
\bibitem{prop}
 M. Agishtein and A. Migdal, Mod. Phys. Lett. A7 (1992) 1039.\\
 J. Ambj\o rn and J. Jurkiewicz, Phys. Lett. B278 (1992) 42.
\bibitem{lots}
  B. Brugmann and E. Marinari, Phys. Rev. Lett. 70 (1993) 1908.\\
 B. V. de Bakker and J. Smit, Phys. Lett. B334 (1994) 304.\\
 S. Catterall, J. Kogut and R. Renken, Phys. Lett. B328 (1994) 277.\\
 J. Ambj\o rn and J. Jurkiewicz, Nucl. Phys. B451 (1995) 643. 
\bibitem{japanese}{\it Singular Vertices in the Strong Coupling Phase of
 Four-Dimensional Simplicial Gravity}, T. Hotta, T. Izubuchi and J.
 Nishimura, UT-724, DPNU-95-29, hep-lat/9511023.
\bibitem{simon} S. Catterall, Computer Physics Comm. 87 (1995) 409.
\bibitem{gross} M. Gross and S. Varsted, Nucl. Phys. B378 (1992) 367.
\bibitem{aergod} J. Ambj\o rn and J. Jurkiewicz, Phys. Lett. B345 (1995)
435.
\bibitem{bergod} B. de Bakker, Phys. Lett. B348 (1995) 35.
 \bibitem{ball}{\it Entropy of random coverings and 4d quantum gravity}
C. Bartocci, U. Bruzzo, M. Carfora and A. Marzuoli, hep-th/9412097.
\bibitem{3d}  S. M. Catterall, J. B. Kogut, and R. L. Renken, Phys. Lett.
 B342 (1995) 53.
\bibitem{usbound} S. M. Catterall, J. B. Kogut, and R. L. Renken, Phys.
Rev.
 Lett. 72 (1994) 4062.
\bibitem{ambbound}J. Ambj\o rn and J. Jurkiewicz, Phys. Lett. B335 (1994)
355.
\bibitem{enzbound} B. Brugmann and E. Marinari, Phys. Lett. B349 (1995) 35. 
\bibitem{baby} 
{\it Baby Universes in 4d dynamical triangulation}, S. Catterall, J. Kogut,
R. Renken and G. Thorleifsson, hep-lat/9509004, Phys. Lett. B in press.

\end{thebibliography}
\end{document}